\documentclass{article}%
\usepackage{graphicx}
\usepackage{amsmath}
\usepackage{amsfonts}
\usepackage{amssymb}
\usepackage{float}%
\providecommand{\U}[1]{\protect\rule{.1in}{.1in}}
\begin{document}

\begin{center}
\bigskip{\LARGE Lorentz-Violating Photon Decay into Neutrinos 

and Constraints from PeV Photon Stability}

{\Huge \bigskip}

{\Huge \bigskip}

\textbf{Z. Kepuladze}\footnote{zurab.kepuladze.1@iliauni.edu.ge,
zkepuladze@yahoo.com}

\bigskip

\textit{Institute of Theoretical Physics, ISU, 0162 Tbilisi,
Georgia\ \vspace{0pt}\\[0pt]}

\textit{Andronikashvili} \textit{Institute of Physics, TSU, 0177
Tbilisi, Georgia\ }
\end{center}

\bigskip

\begin{abstract}
We study the vacuum decay of a Lorentz-violating photon into a
neutrino-antineutrino pair. Lorentz-violating corrections to the photon
dispersion relation are parametrized through an effective invariant mass
$m_{\mathrm{eff}}^{2}=k_{\alpha}k^{\alpha}$. This makes the otherwise
forbidden decay $\gamma\to\nu\bar{\nu}$ kinematically allowed. The process
proceeds through the Standard Model one-loop neutrino electromagnetic vertex
and is therefore strongly suppressed. Using the low-$q^{2}$ anapole form
factor, we derive the decay rate and apply it to TeV and PeV photons. We find
that below the electron-positron threshold the neutrino channel is open but
generally too slow to provide stronger constraints than existing bounds. Above
the threshold, $\gamma\to e^{+}e^{-}$ dominates unless the relative
photon-electron LIV parameter closes this channel. In that case, the neutrino
decay gives an independent constraint on photon-neutrino relative LIV parameters.

\end{abstract}
\bigskip

\section{Introduction}

The detection of photons with energies in the PeV range by the Large High
Altitude Air Shower Observatory (LHAASO) provides a powerful probe of possible
Lorentz-invariance violation (LIV) in the photon sector \cite{LHAASO-PeV}.
Such photons propagate over astrophysical distances, for example from Galactic
PeVatron candidates such as the Crab Nebula ($\sim6.5\times10^{3}$ light-years
away) and the Cygnus region, including the Cygnus OB2 association
($\sim5\times10^{3}$--$6\times10^{3}$ light-years away). Therefore even very
small departures from the standard relativistic dispersion relation can have
observable consequences.

The most commonly discussed consequence of a superluminal photon dispersion is
vacuum photon decay into an electron--positron pair,
\[
\gamma\rightarrow e^{+}e^{-}%
\]
This process gives strong bounds on the LIV deviations of the maximum
attainable velocities of photons and electrons \cite{Coleman-Glashow,
Gagnon-Moore, KM1, Kostelecky-Russell, Stecker, Duenkel-Niechciol-Risse}. In
general, such bounds constrain the difference
\begin{align}
\left\vert \Delta_{\gamma e}\right\vert  &  \equiv\left\vert \delta_{\gamma
}-\delta_{e}\right\vert <5\cdot10^{-21}-5\cdot10^{-22}\\
\delta_{\gamma,e}  &  =\frac{\Delta c_{\gamma,e}}{c}\nonumber
\end{align}
rather than the photon parameter alone. A complementary channel is provided by
the loop-induced decay
\[
\gamma\rightarrow\nu\bar{\nu}.
\]
In the Lorentz-invariant Standard Model this process is kinematically
forbidden for a real photon in vacuum. If LIV makes the photon effectively
timelike, the decay becomes kinematically allowed. Although the corresponding
amplitude is strongly suppressed, since it is generated at one loop (and at
higher orders as well) through the Standard Model neutrino electromagnetic
vertex \cite{Bernabeu-NCR, Cabral-Rosetti-Anapole, Giunti-EM}, it
kinematically probes a different relative parameter,
\begin{equation}
\Delta_{\gamma\nu}\equiv\delta_{\gamma}-\delta_{\nu}.
\end{equation}
It is therefore not directly equivalent to the usual photon-decay or vacuum
Cherenkov bounds involving the electron sector.

We parametrize the LIV correction to the photon dispersion relation by
writing
\begin{equation}
p_{\mu}p^{\mu}=m_{\mathrm{eff}}^{2},\label{effective_mass_definition}%
\end{equation}
where all LIV effects are collected into the effective invariant mass. For a
single preferred four-vector $n^{\mu}$, with
\begin{equation}
p_{n}\equiv n^{\mu}p_{\mu}%
\end{equation}
we consider the phenomenological expansion
\begin{equation}
m_{\mathrm{eff}}^{2}=\mu p_{n}+\delta p_{n}^{2}+\frac{p_{n}^{3}}{M_{1}}%
+\frac{p_{n}^{4}}{M_{2}^{2}}+\cdots.\label{DR}%
\end{equation}
The vector $n^{\mu}$ defines the preferred direction in the LIV vacuum. For
timelike or spacelike $n^{\mu}$ we take $|n^{\mu}n_{\mu}|=1$, while for a
lightlike direction $n^{\mu}n_{\mu}=0$. More general LIV backgrounds could
involve several preferred vectors or tensor structures, but~(\ref{DR}) is the
simplest choice and is naturally motivated in scenarios where LIV is
associated with a single vector structure, for example in models with
spontaneous Lorentz invariance violation by the vector field \cite{Gauge,
Chk1, Chk2, CHk3}.

The terms proportional to $M_{1}^{-1}$ and $M_{2}^{-2}$ are analogous to the
usual cubic and quartic modifications of the photon dispersion relation,
commonly called linear and quadratic LIV corrections in time-of-flight
studies. Recent analyses of LHAASO observations of GRB 221009A give lower
limits of order
\begin{equation}
M_{1}\geq10^{20}~\mathrm{GeV},\qquad M_{2}\geq7\times10^{11}~\mathrm{GeV}%
\end{equation}
for these energy-dependent effects \cite{LHAASO-LIV, Piran-Ofengeim}. The
precise numerical values depend on the sign of the LIV correction and on the
statistical method used. These limits should not be confused with bounds on a
constant velocity shift $\delta$, since time-of-flight analyses are sensitive
to energy-dependent propagation delays. 

However, additional, and in some cases considerably stronger, limits on
$M_{1,2}$ are obtained from modified pair-production processes \cite{Satunin}
and from photon splitting \cite{GNY, AKS, Satunin1}, rather than from
time-of-flight analyses.

The coefficient $\delta$, on the other hand, corresponds to a constant
maximum-velocity shift. As stated above, existing photon-stability and
vacuum-Cherenkov bounds mainly constrain relative parameters such as
$\delta_{\gamma}-\delta_{e}$. In many phenomenological applications one
chooses the charged-lepton sector to be Lorentz invariant, $\delta_{e}=0$, in
which case these bounds are quoted as limits on the photon parameter itself.
However, this is a convention or a model-dependent assumption rather than a
model-independent statement.

The linear term $\mu p_{n}$ in (\ref{DR}) is included here at the
phenomenological dispersion-relation level. It can be generated from the
operator
\begin{equation}
\mathcal{L}_{\mathrm{ext}}=\mu A_{\alpha}(n^{\nu}p_{\nu})A^{\alpha},
\end{equation}
which is not gauge invariant. While reminiscent, it should be distinguished
from the gauge-invariant CPT-odd Carroll--Field--Jackiw/SME operator
\begin{equation}
\mathcal{L}_{AF}=\frac{1}{2}(k_{AF})_{\alpha}\epsilon^{\alpha\beta\mu\nu
}A_{\beta}F_{\mu\nu},
\end{equation}
which is birefringent and is constrained at the level
\begin{equation}
|k_{AF}|\leq10^{-44}~\mathrm{GeV}%
\end{equation}
by astrophysical and cosmological polarization
measurements\ \cite{Kostelecky-Russell, CFJ, KM2}. These stringent
birefringence limits apply directly only to the CFJ-type operator, not to a
polarization-independent phenomenological mass shift such as the one used in
(\ref{DR}).

As a reference scale, for a PeV photon and a timelike LIV correction with
$\delta\sim10^{-22}$, one obtains
\begin{equation}
m_{\mathrm{eff}}^{2}=\delta E_{\gamma}^{2}=10^{8}~\mathrm{eV}^{2},\qquad
m_{\mathrm{eff}}=10~\mathrm{keV}.
\end{equation}
This is well below the electron--positron threshold $2m_{e}\simeq
1~$\textrm{MeV}, so the tree-level decay $\gamma\rightarrow e^{+}e^{-}$ is
forbidden. It is nevertheless far above the neutrino mass scale, so the
channel $\gamma\rightarrow\nu\bar{\nu}$ is kinematically open.

For comparison, if the present lower limits on $M_{1,2}$, are saturated, the
cubic and quartic corrections for PeV energy photon give corresponding values
for $m_{\mathrm{eff}}$ to be $0.1$ and $1.4$ $\mathrm{GeV}$. These values are
significantly larger than the electron--positron threshold. Therefore, for the
superluminal sign and in the absence of cancellations with the electron-sector
LIV parameter, the electron--positron decay channel should dominate. This suggests that photon-stability arguments may lead to stronger constraints on these mass scales than time-of-flight analyses alone. Future observations at higher energies could further improve these limits.

In what follows we compute the decay rate for a general effective photon mass
and then analyze several specific LIV dispersion scenarios in order to extract
possible constraints on the corresponding parameters.

\section{Neutrino electromagnetic coupling}

Standard Model neutrinos do not couple directly to the photon at tree level.
However, an effective electromagnetic coupling is generated through weak
one-loop diagrams. The concepts of the neutrino charge radius and anapole
moment have been discussed in many works \cite{Bernabeu-NCR, Giunti-EM}. In
particular, the neutrino charge radius and anapole moment are not independent
quantities. At lowest order, the effective photon-neutrino vertex is generated
by charged weak-current loop diagrams involving the charged lepton and the $W$
boson \cite{Cabral-Rosetti-Anapole}.

Following \cite{Cabral-Rosetti-Anapole}, the effective vertex for the
electron-neutrino channel may be written as
\begin{equation}
M_{\alpha}=2eF_{D}(k^{2})\gamma_{\alpha}P_{L},
\end{equation}
where
\begin{equation}
k^{2}=k_{\mu}k^{\mu},\qquad P_{L}=\frac{1-\gamma_{5}}{2}.
\end{equation}
Here $e$ is the electromagnetic coupling, $k_{\mu}$ is the photon
four-momentum, which is not necessarily on shell, and $F_{D}(k^{2})$ is the
Dirac form factor. For small $k^{2}$, which means $k^{2}\ll m_{e}^{2}$, one
may expand
\begin{equation}
F_{D}(k^{2})\simeq k^{2}\mathrm{a}_{\nu_{e}},
\end{equation}
where $\mathrm{a}_{\nu_{e}}$ is the above mentioned anapole moment
\begin{equation}
\mathrm{a}_{\nu_{e}}=\frac{G_{F}}{24\sqrt{2}\pi^{2}}\left[  3-2\ln\left(
\frac{m_{e}^{2}}{M_{W}^{2}}\right)  \right]  . \label{anapol}%
\end{equation}
Here $G_{F}$ is the Fermi constant and $M_{W}$ is the $W$-boson mass.

The same expression applies to the muon- and tau-neutrino channels after the
replacement $m_{e}\rightarrow m_{\mu,\tau}$. Thus, more generally,
\begin{equation}
\mathrm{a}_{\nu_{\ell}}=\frac{G_{F}}{24\sqrt{2}\pi^{2}}\left[  3-2\ln\left(
\frac{m_{\ell}^{2}}{M_{W}^{2}}\right)  \right]  ,\qquad\ell=e,\mu,\tau.
\label{anapole}%
\end{equation}
and their numerical values are
\begin{align}
\mathrm{a}_{\nu_{e}}  &  =6.8\times10^{-34}~\mathrm{cm}^{2}=1.75\times
10^{-24}~\mathrm{eV}^{-2}\\
\qquad\mathrm{a}_{\nu_{\mu}}  &  =4.0\times10^{-34}~\mathrm{cm}^{2}%
=1.03\times10^{-24}~\mathrm{eV}^{-2}\\
\mathrm{a}_{\nu_{\tau}}  &  =2.5\times10^{-34}~\mathrm{cm}^{2}=6.42\times
10^{-25}~\mathrm{eV}^{-2}.
\end{align}

In the present problem the "photon virtuality" is fixed by the LIV effective
mass, $k^{2}=m_{\mathrm{eff}}^{2}$. Therefore, for the electron-neutrino
channel, the effective vertex becomes
\begin{equation}
M_{\alpha}=2em_{\mathrm{eff}}^{2}\mathrm{a}_{\nu_{e}}\gamma_{\alpha}P_{L}.
\label{electron}%
\end{equation}
For a general neutrino flavor one has correspondingly
\begin{equation}
M_{\alpha}^{(\ell)}=2em_{\mathrm{eff}}^{2}\mathrm{a}_{\nu_{\ell}}%
\gamma_{\alpha}P_{L},\qquad\ell=e,\mu,\tau. \label{general}%
\end{equation}

As an illustration, for $m_{\mathrm{eff}}=10~\mathrm{keV}$ one has%
\begin{equation}
F_{D}(m_{\mathrm{eff}}^{2})=m_{\mathrm{eff}}^{2}\mathrm{a}_{\nu_{e}}%
\simeq1.75\times10^{-16}.
\end{equation}

\section{Photon decay rate}

The form of the effective vertices in \ref{electron} and \ref{general} shows
that this process is technically very similar to the decay of a massive vector
boson into a neutrino pair. This analogy can be used as a useful shortcut. For
one neutrino flavor $\nu_{\ell}$, the matrix element is
\begin{equation}
\mathcal{M}_{\ell}=2em_{\mathrm{eff}}^{2}\mathrm{a}_{\nu_{\ell}}\xi_{\mu}%
\bar{u}(p)\gamma^{\mu}P_{L}v(q),
\end{equation}
where $\xi_{\mu}$ is the photon polarization vector, while $p_{\mu}$ and
$q_{\mu}$ are the neutrino and antineutrino four-momenta. Momentum
conservation gives
\begin{equation}
k_{\mu}=p_{\mu}+q_{\mu}.
\end{equation}
The squared matrix element is therefore%
\begin{equation}
\left\vert \mathcal{M}_{\ell}\right\vert ^{2}=4e^{2}m_{\mathrm{eff}}%
^{4}\mathrm{a}_{\nu_{\ell}}^{2}\xi_{\mu}\xi_{\nu}Tr[(\not p+m_{\nu_{l}}%
)\gamma^{\mu}P_{L}\left(  \not q-m_{\nu_{l}}\right)  \gamma^{\nu}P_{L}].
\end{equation}
In the regime considered here the effective photon mass is much larger than
the physical neutrino masses. We therefore neglect $m_{\nu_{\ell}}$ in the
phase-space integration and in the trace. This gives
\begin{equation}
|\mathcal{M}_{\ell}|^{2}=4e^{2}m_{\mathrm{eff}}^{4}\mathrm{a}_{\nu_{\ell}}%
^{2}\xi_{\mu}\xi_{\nu}\left[  4p^{\mu}q^{\nu}-2g^{\mu\nu}(p_{\lambda
}q^{\lambda})\right]  .
\end{equation}
The antisymmetric part of the trace does not contribute. We now use
energy-momentum conservation and average over the final-state directions. For
massless final neutrinos,
\begin{equation}
p_{\lambda}q^{\lambda}=\frac{m_{\mathrm{eff}}^{2}}{2},
\end{equation}
and
\begin{equation}
p^{\mu}q^{\nu}\rightarrow\frac{1}{12}\left(  m_{\mathrm{eff}}^{2}g^{\mu\nu
}+2k^{\mu}k^{\nu}\right)  .
\end{equation}
This replacement allows us to avoid specifying the details of the photon
polarization structure in a particular LIV model. We only assume that the
external photon has two physical transverse polarizations satisfying
\begin{equation}
\xi_{\mu}\xi^{\mu}=-1,\qquad k_{\mu}\xi^{\mu}=0.
\end{equation}
Using these relations, the squared matrix element becomes%
\[
|\mathcal{M}_{\ell}|^{2}=4e^{2}m_{\mathrm{eff}}^{4}\mathrm{a}_{\nu_{\ell}}%
^{2}\xi_{\mu}\xi_{\nu}\left[  \frac{1}{3}\left(  m_{\mathrm{eff}}^{2}g^{\mu
\nu}+2k^{\mu}k^{\nu}\right)  -g^{\mu\nu}m_{\mathrm{eff}}^{2}\right]
\]
which simplifies to
\begin{equation}
|\mathcal{M}_{\ell}|^{2}=\frac{8}{3}e^{2}m_{\mathrm{eff}}^{6}\mathrm{a}%
_{\nu_{\ell}}^{2}\label{m_element}.
\end{equation}

Having~(\ref{m_element}) allows to define the decay rate
\begin{equation}
\Gamma_{\ell}=\frac{1}{2k_{0}}\int\frac{d^{3}p}{(2\pi)^{3}2p_{0}}\frac{d^{3}%
q}{(2\pi)^{3}2q_{0}}(2\pi)^{4}\delta^{(4)}(k-p-q)|\mathcal{M}_{\ell}|^{2}.
\end{equation}
Equivalently,
\begin{equation}
\Gamma_{\ell}=\frac{|\mathcal{M}_{\ell}|^{2}}{8\pi^{2}k_{0}}\int\frac{d^{3}%
p}{2p_{0}}\frac{d^{3}q}{2q_{0}}\delta^{(4)}(k-p-q).
\end{equation}
For two massless final particles phase space integration gives
\begin{equation}
\int\frac{d^{3}p}{2p_{0}}\frac{d^{3}q}{2q_{0}}\delta^{(4)}(k-p-q)=\frac{\pi
}{2}.
\end{equation}
Thus,
\begin{equation}
\Gamma_{\ell}=\frac{|\mathcal{M}_{\ell}|^{2}}{16\pi k_{0}}.
\end{equation}
Using~(\ref{m_element}), we finally obtain
\begin{equation}
\Gamma_{\ell}=\frac{e^{2}m_{\mathrm{eff}}^{6}\mathrm{a}_{\nu_{\ell}}^{2}}{6\pi
k_{0}}.\label{decay_L}%
\end{equation}
This expression is written directly in terms of the photon energy $k_{0}$ in
the preferred frame.

Summing over the three neutrino flavors gives
\begin{equation}
\Gamma_{\gamma}=\sum_{\ell=e,\mu,\tau}\Gamma_{\ell}=\frac{e^{2}m_{\mathrm{eff}%
}^{6}}{6\pi k_{0}}\sum_{\ell=e,\mu,\tau}^{2}\mathrm{a}_{\nu_{\ell}}^{2}.
\label{decay total}%
\end{equation}
Equivalently, using $e^{2}=4\pi\alpha$,
\begin{equation}
\Gamma_{\gamma}=\frac{2\alpha}{3}\frac{m_{\mathrm{eff}}^{6}}{k_{0}}\sum
_{\ell=e,\mu,\tau}\mathrm{a}_{\nu_{\ell}}^{2}. \label{decay total1}%
\end{equation}

\section{Consideration and conclusion}

Since we now have the photon decay rate into neutrinos, we can apply it to
astrophysical propagation. Let us assume, as a benchmark, that the photon
lifetime is $\tau_{\gamma}=1000~\mathrm{yr}.$ This corresponds to a decay
rate
\[
\Gamma_{\gamma}=\frac{1}{\tau_{\gamma}}\simeq2.1\times10^{-26}~\mathrm{eV}%
\simeq2.1\times10^{-35}~\text{\textrm{GeV}}.
\]
Such a lifetime would lead to a significant attenuation of photons traveling
over distances of several thousand light-years. Using the (\ref{decay total1}%
), one finds
\begin{equation}
m_{\mathrm{eff}}\simeq3.1~\mathrm{MeV}\left(  \frac{E_{\gamma}}{1~\mathrm{PeV}%
}\right)  ^{1/6}\left(  \frac{1000~\mathrm{yr}}{\tau_{\gamma}}\right)  ^{1/6}.
\label{mef}%
\end{equation}

For PeV photons this gives $m_{\mathrm{eff}}\simeq3.1~\mathrm{MeV},$ whereas
for TeV photons it gives $m_{\mathrm{eff}}\simeq0.99~\mathrm{MeV}.$ The MeV
scale for $m_{\mathrm{eff}}$ is already beyond the strict low-$k^{2}$ range of
the anapole approximation in the (\ref{anapol}) for the electron-neutrino
channel. However, if the (\ref{anapol}) is corrected near the electron
threshold, the correction can only enter as a finite threshold function
multiplying the same gauge-invariant $k^{2}$ structure. Schematically,
\begin{equation}
\Delta a_{\nu_{e}}\sim\frac{G_{F}}{24\sqrt{2}\pi^{2}}\Phi\left(
\frac{m_{\mathrm{eff}}^{2}}{m_{e}^{2}}\right)  ,
\end{equation}
where $\Phi(0)=0$. Away from the threshold $m_{\mathrm{eff}}=2m_{e}$, and for
$m_{\mathrm{eff}}<2m_{e}$, the (\ref{anapol}) should still give the correct
order of magnitude for the electron-neutrino anapole contribution. Near the
threshold an enhancement of the form factor may occur, but this cannot change
the qualitative conclusion by many orders of magnitude.

For $m_{\mathrm{eff}}>2m_{e},$ the decay $\gamma\rightarrow e^{+}e^{-}$ opens
and dominates over the neutrino channel, since it is not suppressed by the
electroweak one-loop vertex with heavy $W$-bosons inside. Therefore, in the
usual case where the electron-positron channel is kinematically allowed, the
corresponding threshold constraint is stronger than the neutrino-decay
constraint. Also, changing the benchmark lifetime by one order of magnitude
does not significantly change the value of $m_{\mathrm{eff}}$, because of the
weak sixth-root dependence in the (\ref{mef}).

The bottom line is the following. If $m_{\mathrm{eff}}$ is decreased below the
MeV scale, the decay rate rapidly decreases and the lifetime increases as
\[
\tau_{\gamma}\propto m_{\mathrm{eff}}^{-6}.
\]
Thus, for smaller values of $m_{\mathrm{eff}}$, propagation of PeV or TeV
photons over several thousand light-years is essentially unaffected by the
neutrino decay channel. For example, decreasing $m_{\mathrm{eff}}$ from
$3~\mathrm{MeV}$ to $1~\mathrm{MeV}$ increases the lifetime by a factor
$3^{6}\simeq7.3\times10^{2}$, which is already close to three orders of
magnitude. For PeV photons, this moves the lifetime from the thousand-year
range to almost the million-year range.

For TeV photons the situation is somewhat cleaner from the point of view of
the neutrino channel. In this case the benchmark value $m_{\mathrm{eff}}%
\simeq1~\mathrm{MeV}$ lies just below the electron-positron threshold,
$2m_{e}\simeq1.022~\mathrm{MeV}$. Therefore the electron-positron channel is
closed, while the neutrino channel is open. Turning on one LIV term at a time
in (\ref{DR}) one obtains the estimates shown in Table~\ref{tab:tev_limits}.

\begin{table}[H]
\centering
\begin{tabular}
[c]{c|c}\hline
Parameter & Estimate for $E_{\gamma}=1~\mathrm{TeV}$, $m_{\mathrm{eff}%
}=1~\mathrm{MeV}$\\\hline
$\mu$ & $\lesssim1~\mathrm{eV}$\\
$\delta$ & $\lesssim10^{-12}$\\
$M_{1}$ & $\gtrsim10^{15}~\mathrm{GeV}$\\
$M_{2}$ & $\gtrsim10^{9}~\mathrm{GeV}$\\\hline
\end{tabular}
\caption{Order-of-magnitude parameter estimates from the neutrino decay
channel for a $1~\mathrm{TeV}$ photon with $m_{\mathrm{eff}}\simeq
1~\mathrm{MeV}$. Each line assumes that only the corresponding LIV term is
present.}%
\label{tab:tev_limits}%
\end{table}

These numbers do not improve existing limits, especially when compared with
the constraints obtained from the electron-positron decay threshold. The
neutrino channel is nevertheless conceptually different, since it probes the
photon-neutrino relative LIV parameter rather than the photon-electron one.

In principle, it is possible to have
\[
m_{\mathrm{eff}}\simeq3~\mathrm{MeV}
\]
and still avoid the decay $\gamma\to e^{+}e^{-}$. This can happen because the
electron-positron threshold depends on the relative difference between the
photon and electron LIV parameters, not on the photon parameter alone. If this
relative photon-electron combination is arranged so that $\gamma\to e^{+}%
e^{-}$ is closed, then the neutrino channel gives an independent constraint on
the photon-neutrino relative LIV parameters. For PeV photons, taking
\[
E_{\gamma}=1~\mathrm{PeV}, \qquad m_{\mathrm{eff}}\simeq3~\mathrm{MeV},
\]
one obtains the estimates shown in Table~\ref{tab:pev_limits}.

\begin{table}[H]
\centering
\begin{tabular}
[c]{c|c}\hline
Parameter & Estimate for $E_{\gamma}=1~\mathrm{PeV}$, $m_{\mathrm{eff}%
}=3~\mathrm{MeV}$\\\hline
$\mu$ & $\lesssim9\times10^{-3}~\mathrm{eV}$\\
$\delta$ & $\lesssim9\times10^{-18}$\\
$M_{1}$ & $\gtrsim1\times10^{23}~\mathrm{GeV}$\\
$M_{2}$ & $\gtrsim3\times10^{14}~\mathrm{GeV}$\\\hline
\end{tabular}
\caption{Order-of-magnitude parameter estimates from the neutrino decay
channel for a $1~\mathrm{PeV}$ photon with $m_{\mathrm{eff}}\simeq
3~\mathrm{MeV}$, assuming that the electron-positron decay channel is
kinematically closed. Each line assumes that only the corresponding LIV term
is present.}%
\label{tab:pev_limits}%
\end{table}

If we apply (\ref{decay total1}) to an extragalactic source, such as GRB
221009A, located at a distance of about $700$ \textrm{Mpc}, and take a
benchmark photon lifetime of $10^{9}$ years, then for photons with energies up
to $18$ \textrm{TeV}, as reported by LHAASO, (\ref{mef}) gives
$m_{\mathrm{eff}}\simeq0.16~\mathrm{MeV}$. LHAASO reported more than 5000
very-high-energy photons from GRB 221009A, with the highest-energy event
reaching approximately $18$ \textrm{TeV }\cite{LHAASO-GRB, LHAASO-GRB1}. The
corresponding limits are shown in Table \ref{tab:grb_limits}. 

\begin{table}[H]
\centering
\begin{tabular}{c|c}
\hline
Parameter &
Estimate for $E_\gamma=18~{\rm TeV}$,
$m_{\rm eff}\simeq0.16~{\rm MeV}$ \\
\hline
$\mu$    & $\lesssim 1\times10^{-3}~{\rm eV}$ \\
$\delta$ & $ \lesssim 8\times10^{-17}$ \\
$M_1$    & $ \gtrsim 2\times10^{20}~{\rm GeV}$ \\
$M_2$    & $ \gtrsim 2\times10^{12}~{\rm GeV}$ \\
\hline
\end{tabular}
\caption{Order-of-magnitude parameter estimates obtained for an
$18~{\rm TeV}$ photon with $m_{\rm eff}\simeq0.16~{\rm MeV}$,
corresponding to the benchmark lifetime $\tau_\gamma=10^9~{\rm yr}$.
Each line assumes that only the corresponding LIV term is present.}
\label{tab:grb_limits}
\end{table}

These bounds are considerably stronger than those obtained for the Galactic
TeV benchmark in Table~\ref{tab:tev_limits}. However, the bounds on $\delta$,
$M_{1}$ and $M_{2}$ still do not improve the strongest existing constraints.
In particular, photon-splitting analyses provide strong limits on cubic and
quartic LIV corrections \cite{GNY, AKS, Satunin1}. The parameter $\mu$, on the
other hand, is not covered by these photon-splitting bounds.

If photons with energies around $100$ \textrm{TeV} are detected from
comparable extragalactic distances, while keeping the same benchmark lifetime,
from the (\ref{decay total1}) we estimate $m_{\mathrm{eff}}\simeq
0.2~\mathrm{MeV}$. The corresponding limits would then be substantially
stronger, as shown in the table below.

\begin{table}[H]
\centering
\begin{tabular}{c|c}
\hline
Parameter &
Estimate for $E_\gamma=100~{\rm TeV}$,
$m_{\rm eff}\simeq0.21~{\rm MeV}$ \\
\hline
$\mu$    & $ \lesssim 5\times10^{-4}~{\rm eV}$ \\
$\delta$ & $ \lesssim 5\times10^{-18}$ \\
$M_1$    & $ \gtrsim 2\times10^{22}~{\rm GeV}$ \\
$M_2$    & $ \gtrsim 5\times10^{13}~{\rm GeV}$ \\
\hline
\end{tabular}
\caption{Order-of-magnitude parameter estimates for a
$100~{\rm TeV}$ photon with $m_{\rm eff}\simeq0.21~{\rm MeV}$,
assuming the benchmark lifetime $\tau_\gamma=10^9~{\rm yr}$.
Each line assumes that only the corresponding LIV term is present.}
\label{tab:100tev_limits}
\end{table}

These estimates do not improve the strongest existing bounds on the constant
velocity parameter $\delta$. However, they show that, if the electron-positron
channel is absent because of relative photon-electron LIV cancellations, the
neutrino channel can still provide an independent restriction on $\mu$,
$M_{1}$, and $M_{2}$. Strictly speaking, for the energy-dependent terms this
restriction applies to the photon-neutrino relative LIV combination. It
coincides with a direct photon-sector bound only if the corresponding neutrino
LIV correction is absent or significantly smaller.

\section*{Acknowledgments}

The author expresses gratitude to Jon Chkareuli for useful discussions.

\end{document}